\documentstyle[aps,prb,multicol,graphicx]{revtex}

\begin{document}
\draft
\widetext

\title{Theory of the Diamagnetism Above the Critical Temperature 
for Cuprates}

\author{J.L. Gonz\'alez and E.V.L. de Mello}

\address{Departamento de F\'{\i}sica,
Universidade Federal Fluminense, Niter\'oi, RJ
24210-340, Brazil  }
\date{Received \today }
\maketitle

\begin{abstract}

Recently experiments  on high critical temperature  superconductors 
have shown that the doping levels and the superconducting gap
are usually not uniform properties but strongly dependent
on their positions inside a given sample. Local superconducting regions 
may develop at the pseudogap temperature ($T^*$) and upon cooling, grow continuously. 
As one of the consequences a large diamagnetic signal above 
the superconducting critical temperature ($T_c$) has been
measured by different groups. Here we  construct a general theory to
a disordered superconductor using
a critical-state  model for the magnetic response 
to  the local superconducting domains  between 
$T^*$ and $T_c$ and show that the resulting diamagnetic signal is 
in agreement  with the experimental results.

\end{abstract}

pacs{74.20.-z,74.25-q,74.25.Ha,74.72-h }

\begin{multicols}{2}
\section{Introduction}
There is an increasing interest on the effects of the
microscopic intrinsic inhomogeneities in
high critical  temperature  superconductors
(HTSC). The fact
that these materials have doping level and superconducting
gap  that, even in some of the best single crystals, vary 
locally inside 
a given sample,  has profound consequences and
a number of unconventional related phenomena are observed\cite{Egami96}
specially in the normal region of the HTSC phase diagram. In particular,
recent magnetic imaging through  a scanning superconducting quantum 
device (SQUID) microscopy has displayed a static  local precursor of the  Meissner
state at temperatures as large as three times the $T_c$ of an underdoped
LSCO film\cite{Iguchi}. Following up SQUID 
magnetization measurements on powder oriented  YBCO and LSCO single
crystals\cite{Lascialfari1,Lascialfari2} have shown a rather high
magnetic response which, due to its large signal and structure, cannot be 
attributed solely to the Ginzburg-Landau (GL) theory of fluctuating
superconducting magnetization\cite{Puech,Sewer}. This is an unconventional 
behavior because low temperature superconductors
exhibit only a much smaller diamagnetic signal above $T_c$
which is believe to be a consequence of the thermal 
fluctuations\cite{Tinkham}. Therefore such strong magnetic 
response was interpreted  as due to the 
fluctuating diamagnetism produced by superconducting islands
nucleated above $T_c$\cite{Lascialfari1,Lascialfari2,Romano}. 
On the other hand, the existence of superconducting 
islands above $T_c$ is a 
consequence of a non-uniform or disordered intrinsic carrier density,
 as pointed out by Ovchinnikov et al\cite{OWK1}.

 The tendency towards the formation of  magnetic domain lines or
stripes was predicted long ago\cite{Zaanen} and verified experimentally
by different groups\cite{Tranquada,Bianconi}. The stripes are formed by
an instability towards charge phase separation  which 
means hole-poor regions of antiferromagnet (AF) insulating  separated by
hole-rich metallic domains. 
Recently, the microscopic intrinsic inhomogeneities  in the charge distribution,
consistent with the presence of charge domains or charge stripes,
has been revealed by neutron diffraction\cite{Buzin00}.
A peak broadening in the atomic pair distribution function was measured
and explained by a local microscopic coexistence of
doped and undoped material\cite{Buzin00}.
Evidences for microscopic phase separation into normal and superconducting
regions was also obtained by muon spin relaxation rate\cite{Uemura}.
Therefore the inclusion of  the intrinsic  charge inhomogeneities or the
tendency of the formation of charge domains
in any calculations for the cuprates properties  may be far 
more important than previously anticipated.  
The main difficulty to incorporate such effect in quantitative
calculations is that the exact form of the doping or hole distribution 
inside a given sample is
not exactly known although it is a consensus that the local doping level becomes more
uniformly distributed in the materials on crossing from the
underdoped into the  overdoped region of the phase diagram\cite{Egami96,Buzin00}.
On the other hand, a considerable 
improvement in the understanding of the doping level and local superconducting gap
functional form has been made by scanning tunneling microscopy/spectroscopy 
(STM/S)  experiments\cite{Iguchi,Fournier,Pan,Davis}. 

Based on these  findings, we have 
recently proposed a two-phase model (undoped and doped regimes present
in a given compound) described
by a bi-modal distribution 
of the holes inside a given compound in
order to model the charge distribution for a  given compound belonging 
to a HTSC family\cite{EEJ}. This distribution resembles a typical 
distribution of a spinodal decomposition of
a binary alloy\cite{Cahn}. A two-phase model which arises from the
fluctuation phase of the superconducting order parameter was considered
previously by Emery et al\cite{Emery}.
The basic ideas of our approach are: The undoped or  hole-poor part of 
the distribution
represents the AF domains and the hole-rich the metallic regions.
The width of the metallic distribution decreases with the sample's average doping 
since,  as mentioned above, the hole distribution  inside 
the compounds of a given family become more homogeneous as the
average doping level or average charge density  increases\cite{Buzin00}. 
Due to the spatially varying local charge density, it 
is also expected a distribution of the local $T_c$'s, instead of  a single
and unique value as in usual metallic superconductors. 
Therefore a given HTSC compound
with an average charge density $n_m$ 
possess distributions of the
charge density $n(r)$, the zero temperature  superconducting gap $\Delta_0(r)$ 
and the superconducting critical  temperature $T_c(r)$,  where the
symbol $(r)$ means a point or small region inside the sample. 
In this scenario we identify the largest  $T_c(r)$  
with the the pseudogap temperature $T^*$ of
a given  compound\cite{TS}.  Doping regions with $n(r) > 0.05$ have a 
metallic behavior with  a decreasing  $T_c(r)$ following a mean field
pattern\cite{EK}. Upon cooling below $T^*$ the superconducting
regions develop at isolated regions (mostly with $n(r) \leq 0.05$) 
as droplets of rain in the
air and,  eventually they percolate at the sample
superconducting critical temperature $T_c$  at which
superconducting long range order is established. This
superconducting percolation scenario for HTSC
has also been discussed previously, but with somewhat different
approaches, by several authors\cite{OWK,Mihailovic}. 
Therefore, only at or  below the percolation temperature  $T_c$
a dissipationless macroscopic electrical  or hole current may 
flow through the sample. 
Based on these ideas, we have used  a bi-modal charge distribution 
in connection with a  mean field calculation  to 
reproduce  the phase diagram of $T^*$  and $T_c$
as function of the average doping level for the $Bi_2Sr_2CaCu_2O_{8+x}$
and $La_{2-x}Sr_xCuO_4$ families\cite{EEJ,prbrap}. In these chemical
formulas $x$ is the doping level which is similar to the
average charge density, i.e., $x=n_m$ for the last  (La series) and
$x=2n_m$ for the former (Bi) family of compounds 

The existence of the
superconducting regions with different $T_c$'s in the material must
manifest itself through a number of observable properties: we can 
number a few like the pseudogap phenomena, the resistivity and
Hall coefficient temperature dependence, and  so on. Here we want to discuss the
normal state magnetization, whose study is the purpose of this paper.
Thus, we develop a general theory of the  magnetization of a inhomogeneous 
superconductor which can be applied to the HTSC, in 
order to explain the recent anomalous magnetization 
measurements\cite{Lascialfari1,Lascialfari2}. In a similar fashion,  
L. Romano\cite{Romano}  introduced a theory for the magnetic
response based on the fluctuations in the local order parameter which
leads to a Kosterlitz-Thouless transition and the fluctuation-induced 
diamagnetism\cite{Sewer}
of superconducting islands above $T_ c$ which reproduced well
the details of the experimental data. Assuming that the superconducting islands
started to be formed at $T^*$, for measurements at $T<T^*$ and
$T$ close but above  $T_c$, some of these islands must be in a
superconducting state and therefore they must yield a 
local superconducting diamagnetic response. 
We show below that the measured  diamagnetic signal
can be understood within this picture
of a superconductor formed by static 
domains with  spatial varying $T_c(r)$s.
Since the local $T_c(r)$'s can be much higher than the sample's 
superconducting $T_c$, we apply the well known critical-state 
model (CSM)\cite{Tinkham} to the magnetization
response under an applied field. 
This approach is valid with the hypothesis of the measured temperature to
be below that of the irreversibility line of the local superconducting 
regions . Some kink of hysteretic
behavior was observed in  YBCO and in LSCO 
samples\cite{Lascialfari1,Lascialfari2} indicating that the
irreversible temperature is above $T_c$. 
We demonstrate here that this simple procedure
is able to explain and reproduce the main features of the precursor diamagnetism 
measured behavior.

\section{The Model}
From the above discussion, since the hole distribution is not uniform,
we can think of a HTSC sample as formed by different
metallic an insulating regions of a few nanometers of magnitude.
As a consequence of such distribution, at a given  temperature  above
$T^*$ there are insulating and metallic regions, below
$T^*$, there  are insulating, metallic and superconducting  
regions which started to condensate in the metallic regions
at $T^*$ and below $T_c$ the long range superconducting order
is achieved by a percolation transition while there still are 
some metallic and insulating regions present in the sample. 
As in Ref.\cite{EEJ}, we model these insulator and 
metallic regions by  a bi-modal  Gamma-type 
distribution with parameters
which are based on the STM/S analysis\cite{Pan} 
and which was used to derive the HTSC phase diagrams.

Now we want to use the above ideas  to discuss the 
magnetic response $M(B)$ of such inhomogeneous 
superconductors to an external magnetic field, 
between $T_c$ and $T^*$. In this range
of temperature, it is clear that only the superconducting 
regions or droplets with their $T_c(r)$'s bigger than $T_c$ 
contribute to the diamagnetic sample's magnetization. 
These  superconducting region contributes to the sample's 
magnetization with a diamagnetic signal which depends on
$B$  and the overall
sample's magnetization will be sum of all these contributions. 

In order to estimate the $M(B)$ we follow the ideas and
the procedures of the CSM to each superconducting droplet.
Upon applying an external magnetic field, a 
critical current ($J_c$) is established
which opposes the field as $J_c(B) = \alpha(T)/B$
according to Ohmer et al\cite{Ohmer}. 
$\alpha(T)$ is a temperature-dependent local constant which we have
taken to be proportional to $(T_c(n(r))-T)$.  Hereafter we
call the hole local density $n(r)$ of a given domain of $n$ and 
the whole sample's average density of $n_m$. 
For simplicity we take these superconducting droplets as cylinders
of radius $R$, which is sufficient small in order to have
a constant charge density $n$ and consequently the
critical temperature $T_c(n)$ is the same 
within such cylinder region (As the temperature decreases, more
droplets appear and the superconducting regions increases 
by aggregation of droplets of different $n$).
The CSM approach leads to the magnetic field dependence of the
magnetization in each small cylinder as\cite{Ohmer}:

\begin{eqnarray}
 M_1(B)&=& -\frac{B}{\mu_0}  \; for \; B\leq B_{c1} \\
 M_2(B)&=& -\frac{B}{\mu_0}+\frac{4B^3}{5\mu_0B^{*2}} -\frac{8B^5}
{15\mu_0B^{*4}} \; ,  \; B_{c1}\leq B\leq B^*\\
 M_3(B)&=& -\frac{B}{\mu_0}-\frac{4B^*}{15\mu_0}(2\frac{B^5}{B^{*5}}-5\frac{B^3}{B^{*3}}
 \nonumber  \\&& -2[\frac{B^2}{B^{*2}}-1]^{5/2}) \; for \;
 B_*\leq B\leq B_{c2}
\end{eqnarray}

In the last expression, $B^*$ is the value of the applied external field 
which produce full penetration inside a cylindrical superconducting droplet, 
and it depends on its size  as $B^*=\sqrt{2\alpha(T)\mu_0R}$. 
$\mu_0$ is the vacumn permissivity. The 
dependence of the local critical temperatures 
$T_c(n)$ is taken to be  that of the pseudogap
temperature of a given compound with its average 
charge density, namely $T^*(n_m)$. Thus, $T_c(n)$ is a
function which has its maximum at $n=0.05$ and
decreases monotonously, that is, we take the 
local superconducting critical temperatures
as a linear function of $n$, $T_c(n)=T_0-b*(n-n_c)$. $T_0$ is
the highest pseudogap temperature near the onset
of superconductivity, i.e., $n_c=0.05$, and $b$ is chosen 
in order to $T_c(0.3)=0$. Notice that, since $T_c(n)$ is 
a linear decreasing function of $n$, the droplets
with lower values of $n$ (just above $n_c$) are more robust to an applied field
in the sense that its superconducting state is not
easily destroyed by the external field.


 Since $T_c(n)$ is constant inside a superconducting 
cylindrical droplet, the critical fields ($B_{c1}$ and $B_{c2}$) 
inside the droplets will have their temperature dependence given by the
GL theory, e. g., $B_{c1}(T)=B_{c1}(0)[(1-T/T_c(n)]$ 
and $B_{c2}(T)=B_{c2}(0)[(1-T/T_c(n)]$. Taking into account 
the dependence of $T_c(n)$ on $n$, we arrive at the
expressions for the critical fields $B_{c1}(T,n)$ and
$B_{c2}(T,n)$. A similar functional form is supposed for $B^*$ due to
the $\alpha(T)$ temperature dependence. Thus,

\begin{equation}
B_{c1}(T,n)=B_{c1}(0)[1-\frac{T}{T_0-b(n-n_c)}]
\end{equation}				     
\begin{equation}
B_{c2}(T,n)=B_{c2}(0)[1-\frac{T}{T_0-b(n-n_c)}]
\end{equation}				     
\begin{equation}
B^*(T,n)=B^*(0)[1-\frac{T}{T_0-b(n-n_c)}].
\end{equation}				     

When a given sample is submitted to an applied external
magnetic field $B$, the
superconducting droplets with carrier concentration 
$n$ for which the applied field is higher than their
second critical field $B_{c2}(T,n)=B_{c2}(0)[(1-T/(T_0-b*(n-n_c)]$, 
do not contribute to the sample magnetization. 
This condition is verified for droplets with $n>n_{max}$, where 
$n_{max}(B_{c2})=n_c+T0/b-(T/b)/[1-B/B_{c2}(0)]$ is obtained
inverting Eq.5.  Since $T_c(n)$ is a decreasing function of $n$, 
only the droplets with $n$ bigger than 
$n_{max}$  do not contribute 
to the sample's magnetization because their superconductivity
is destroyed by the field $B$. Thus we expect that 

\begin{equation}
 M(T,B)= \int^{n_{max}(B_{c2})}_{n_c} P(n)M(B,T,n)dn.
\end{equation}

Where $P(n)$ is the distribution function  for the 
local hole doping level at the many clusters inside a given 
HTCS inferred in Ref.\cite{EEJ}.
However, in the context of the CSM, depending on the intensity
of the applied field, there are  different
possibilities  in which each domain contributes to $M(B)$.
In the low field regime the superconducting clusters will 
contribute to the magnetization of the sample in three
forms: there are some clusters, which are not  
penetrated by the field $B$, that is  $B \le B_{c1}(T,n)$ 
and they contribute to the
magnetization with perfect diamagnetism (Eq.1). 
These clusters have their carrier concentration in the
interval, $n<nc +T0/b-(T/b)/[1-B/B_{c1}(0)]$. The second 
group of clusters have their $B_{c1}(T,n)$ lower than the applied
field but $B$ is also lower than $B^*(n,T)$. This group is 
partially penetrated by the field and they contribute to $M(B)$
according Eq.2. These domains have their carrier 
concentration in the  interval 
$n_c+T_0/b-(T/b)/[1-B/B_{c1}(0)]<n< n_c+T_0/b-(T/b)/[1-B/B^*(0)]$. 
Lastly, there are some superconducting granules for which
the applied field is higher than 
$B^*(T,n)$ but also lower than $B_{c2}(T,n)$. 
These domains contribute to the magnetization according Eq.3. 
Therefore, for a sufficient low applied field, the general
expression for the $M(B)$ is given by:

\begin{eqnarray}
 M(T,B)&=& \int^{n_{max}(B_{c1})}_{nm} P(n)M_ 1(B,T,n)dn
 \nonumber \\&& + \int^{n_{max}(B^*)}_{n_{max}(B_{c1})} P(n)M_ 2(B,T,n)dn
 \nonumber \\&& + \int^{n_{max}(B_{c2})}_{n_{max}(B^*)} P(n)M_ 3(B,T,n)dn
\end{eqnarray}

Upon increasing the  applied field the number of droplets that are 
not penetrated by the field decreases and for fields
higher than $B_{c1}(n_m,T)$ there are not any granules contributing with 
perfect diamagnetism. Notice that $Bc1(n_m,T)$ is the
maximum first critical field that a superconducting region can 
achieve.  For $B\geq B_{c1}(n_m,T)$, all the droplets
are partially or totally penetrated by the field and they 
contribute to the sample's magnetization according Eqs.1-3.
\begin{eqnarray}
 M(T,B)&=& \int^{n_{max}(B^*)}_{nm} P(n)M_ 2(B,T,n)dn
\nonumber \\&& + \int^{n_{max}(B_{c2})}_{n_{max}(B^*)} P(n)M_ 3(B,T,n)dn
  \end{eqnarray}
Finally for fields higher than $B^*(0)[(1-T/(T0-b(n_m-n_c)]$, 
all the superconducting granules of the sample are
totally penetrated by the field and their contribution 
to the magnetization is
\begin{equation}
 M(T,B)= \int^{n_{max}(B_{c2})}_{nm} P(n)M_ 3(B,T,n)dn .
 \end{equation}
This last three equations express the total $M(B)$ for any value of the 
applied field.

In order to obtain a reliable value of $M(B)$ and to compare 
with the experimental results, we have incorporated the fluctuation
magnetizations induced by the superconducting order parameter,
an effect which should be always present, regardless whether the
superconductor is more or less inhomogeneous. 
As noted in reference\cite{GBCT},
for superconducting droplets with a homogeneous order parameter and with
dimensions $d$ approximately equal to the coherent length $\xi(T)$, the
Ginzburg-Landau model provides an exact solution for $M_{fluct}(B)$. 
Here we use a simplified "zero-dimensional" for superconducting
clusters of radius $d$ smaller or near  the coherence length $\xi(T)$,
namely\cite{Lascialfari1}:
\begin{equation}
M_{fluct}(T,B)=  - \frac{2/5k_B(\pi \xi Td)^2B}{\Phi_0^2(T/T_c-1)+(\pi \xi Bd)^2/5}
\end{equation}
where  $k_B$ is the Boltzmann constant, $\Phi_0$  is the quantum flux 
and $d\approx \xi \propto (T-T_c)/T_c$. This last expression yields a
linear $M_{fluct}(B)$ dependence for low fields and it has  been incorporated
in our calculations. The specific results for $M_{fluct}$ are shown in
the insets of Figs.1-2.

Therefore Eqs.7-11 furnish the complete $M(B,T)$ curve
for a HTSC at temperatures which  $T^*>T>T_c$. Our numerical results
will be compared with the experimental results in 
the following section.

\section{Results and Discussion}
  
The above theory was developed  to model the  
measured magnetization curves of the 
$ La_{1-x}Sr_xCuO_4$ family of compounds. The major difficulty
is that the measurements of the pseudogap temperature curve 
for any family seems to vary depending on the technique used\cite{TS}.
In fact there are differences even in the $T_c$ measured 
values for the same HTSC compound when
reported by different groups, but such differences are small 
compared with those for the $T^*$ measurements. 

In the calculations we used values taken from the local precursor diamagnetic
signal\cite{Iguchi} and Nernst effect\cite{Xu} measurements which display some
local precursor order near $T\approx 100$K. This value is considerably lower  than
the values of $T^*$ from transport measurements\cite{TS}. A possible
explanation for this discrepancy is that
the transport measurements detect the onset of pair formation while for
the magnetic response a large density of pairs must be present in order to
display a measurable local response.
Thus, we have taken  
$T_0= 80$K, $b = 320$K for the carrier concentration linear dependence of the 
droplet's critical temperature $T_c(n)$. In fact $T_c(n)$ is not
necessarily linear but this is an approximation which fits reasonably
well the experimental data but which may be improved in the 
future as new experimental data becomes available. 
The antiferromagnetic cutoff 
for the  carrier concentration is $n_c = 0.05$ and the superconducting
cutoff is $n_m=0.3$. The probability $P(n)$ is also not well known
but some of its feature was inferred  from STM measurements and 
the phase diagram, as discussed in Ref.\cite{EEJ}.

In Fig.1 we plot the results of our model with the parameter  which
corresponds to a $n_m=0.2$. According to our model this concentration 
corresponds to an LSCO overdoped sample with a critical temperature 
of $T_c= 26.86$K. In the calculations we have 
used $B_{c1}(0)=0.01$T, $B_{c2}(0)=20$T and  $B^*(0)=0.025$T. 
The value of $B_{c2}$ is derived from experimental values\cite{Ando}.
The values of $B_{c1}(0)$ and $B^*(0)$ are not found in the literature,
however one can make a fair estimation of the ratio $B_{c2}/B_{c1}$
through the GL parameter $\kappa$ which is  about 100
for some YBCO and LSCO compounds\cite{Poole}.
The inset shows that for $T = 27.8$K there
is a low contribution at low fields from the superconducting
island in the sample  while
at high fields only a linear contribution comes from the diamagnetic
fluctuation signal. At $T = 28$K there is
no more contribution from the superconducting droplets and all the
diamagnetic signal is only due to the magnetization fluctuations. Note
that the T = 28 K is the most week diamagnetic signal which
appears in the Fig.1. The inset demonstrate that, at this 
temperatures or above, only the fluctuating magnetization 
is important.

The qualitative features of the measurements are entirely
reproduced and are simply explained by our model; at low
fields the perfect diamagnetism is expected for droplets
for which the fields are lower
than their $B_{c1}$. We expect $B_{c1}$ to be weak 
because  the superconducting regions
formed above $T_c$ are small  and isolated. 
By the same token, the droplets penetrating
field $B^*$ should not be very strong what decreases
rapidly the overall diamagnetic signal for field much
weaker than $B_{c2}$. As the applied field increases, the 
magnetic response dies off and is reduced to the 
fluctuations. This is the reason why $M(B)$
has a minimum at very low applied fields. 
In Fig.1
we show this up-turn field near $B_{up}=0.0001$T which agrees
with the experimental values\cite{Lascialfari2}. It is 
worthwhile to mention that previously estimation for the
up-turn field considering only the Lawrence-Doniach fluctuations\cite{Puech}
in a layered superconductor\cite{Lascialfari1} yields expected values near
$B_{up}=10$T. These figures bring out the importance of the
CSM applied to the  superconducting islands above 
$T_c$ to explain the  experimental results.

\begin{figure}
\includegraphics[width=9.5cm]{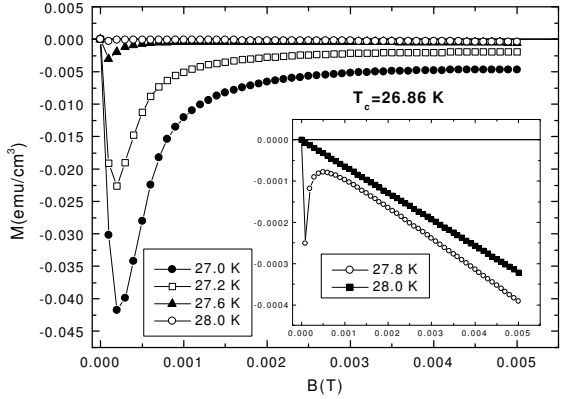}
\caption{Magnetization for parameters appropriated to the overdoped LSCO  as
calculated from Eqs. 7-11. The inset shows the change in the magnetization
behavior as the anomalous contribution vanishes when 
the temperature is increased. This result is to be compared
with the measurements from ref.4.}
\end{figure}
In order to study the effect of the local charge inhomogeneities, we have also performed
similar calculations with parameters appropriate to an underdoped
compound. It is well known that, within the same family, overdoped 
samples have a more homogeneous charge distribution  than 
the underdoped ones\cite{Buzin00}. 
Thus, we have chosen a compound with almost the same critical 
temperature $T_c$ than that of the overdoped sample (shown in Fig.1) to single out 
the effect of the charge inhomogeneities. In this simulation we used 
a carrier concentration of average concentration of $0.11$ with its appropriate 
distribution $P(n)$
and with a critical temperature  of $29.9$K. We
have chosen the same field parameters, e.g. $B_{c1}(0)=0.01$T, $B_{c2}(0)=20$T 
and  $B^*(0)=0.025$T.
The results from our simulations are shown in Fig.2. In this case 
the overall diamagnetic signal persists for measurement 
temperatures higher than those for the overdoped sample. 
The $B_{up}$ field is dislocated to higher values but it is
still much lower than the homogeneous Lawrence-Doniach\cite{Puech}  
fluctuating field.
This is a consequence of the fact that a underdoped sample has a more
inhomogeneous charge distribution than the overdoped compounds 
possessing regions where the field easily penetrates (with local $n\approx 0.25$)
and others where are more robust to field penetration (with local $n\approx 0.1$
for instance). This fact 
is taken into account through the distribution parameters used in our 
simulation. The results for an underdoped compound are shown in Fig. 2 
where can be also noticed in the inset that as the temperature 
approaches 37K, the diamagnetic contribution of the superconducting droplets 
decreases and for measurements temperature higher than approximately 37K 
there is only the fluctuations contribution.

\begin{figure}
\includegraphics[width=9.5cm]{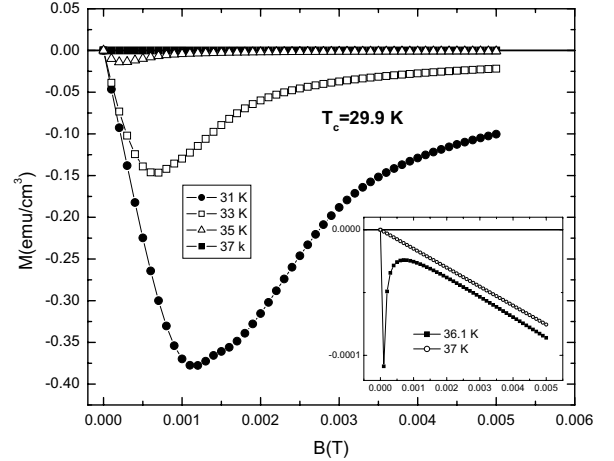}
\caption{Magnetization for parameters appropriated to the underdoped LSCO  as
calculated from Eqs. 7-11. The inset shows the change in the magnetization
behavior as the anomalous contribution vanishes when 
the temperature is increased. This result is to be compared
with the measurements from ref.4.}
\end{figure}
It is important to mention that, although our calculations are very much in
qualitative agreement with the magnetic imaging results\cite{Iguchi} and 
with the powder oriented SQUID magnetic measurements\cite{Lascialfari1,Lascialfari2},
there are other experiments which have not detected a $B_{up}$ field\cite{Carballeira}
and have interpreted their results as solely due to the GL
fluctuating diamagnetic theory\cite{Sewer}. However an inspection in
their curves indicates that they have not looked in detail at
the low field region. The absence of a $B_{up}$ field was also found
on an optimally doped powder oriented YBCO 
compound\cite{Lascialfari1} whose curves followed the pure
fluctuating magnetic response. In this case, the possible interpretation for a
pure fluctuating diamagnetism signal, according to our
theory, is that these samples must have
a very good degree of charge homogeneity. This interpretation can  be
experimentally tested because these compounds must
have $T^*$ almost equal or equal $T_c$, as may be the case\cite{TS}. 

Another point which hinders a good 
quantitative agreement with the experimental results 
is that the technique used seems also
to be very important in probing the details of the magnetization above 
$T_c$; while the scanning microscopy results on a LSCO film indicate the
presence of a visible diamagnetic signal at temperatures as
large as three times the sample's $T_c$\cite{Iguchi}, the
SQUID magnetization measurements detected a diamagnetic signal 
on oriented powder
only around $10\%$ above $T_c$\cite{Lascialfari1,Lascialfari2}.

It is important to emphasize that, in our calculations of $M(B,T)$, 
the islands that contribute to the diamagnetic signal are in
the critical state, and the measured temperature $T$ is
below their local $T_c(r)$.  In the other hand, the Romano's calculations
yield an upturn field which is almost independent of the temperature
while in Fig. 2 we can see clearly that it decreases with the
applied temperature. Future experiments may distinguish which 
mechanism is more important or if some compounds exhibit one
type and others exhibit the other.

\section{Conclusion}
 We have constructed a general theory of the diamagnetism 
applicable to a 
disordered superconductor with a distribution of regions with different 
local superconducting $T_c$'s.
We have shown that the procedure 
reproduces the qualitative features
of the unusual diamagnetic signal above $T_c$ measured for several
HTSC samples. This was done applying a
CSM and the well known  magnetic
effects for fields between $B_{c1}$  and $B_{c2}$ to the superconducting
regions formed at temperatures below $T^*$  
of an inhomogeneous HTSC compound. The quantitative results can be improved when
experiments provide better
estimates of $T^*$ and $B_{c1}$.

Our results demonstrated that, as concluded also from others different
calculations\cite{Lascialfari2,Romano}, the measured normal state
magnetization curves, the $B_{up}$ fields and the STM 
magnetic imaging results may be interpreted through 
the formation of static superconducting islands at temperatures
above the sample's $T_c$.
\section{Acknowledgments}
We want to thank prof. A. Rigamonti for discussions that led
to this work and for providing us with the experimental results prior to
their publications. Partial financial aid from CNPq and FAPERJ
is gratefully acknowledged.

\end{multicols}
\end{document}